\documentclass[pra,aps,twocolumn,showpacs,amsmath,amssymb,superscriptaddress]{revtex4-1}

\usepackage{graphicx}
\usepackage{color}
\usepackage{dcolumn}
\usepackage{bm}
\usepackage{epsf}
\usepackage{amsmath,amstext,amssymb}
\usepackage{enumitem}
\usepackage{hyperref}
\usepackage[ansinew]{inputenc}
\usepackage{epstopdf}

\newcommand{\be}{\begin{equation}}
\newcommand{\ee}{\end{equation}}
\newcommand{\bea}{\begin{eqnarray}}
\newcommand{\eea}{\end{eqnarray}}

\begin{document}

\title{Microscopy with a Deterministic Single Ion Source}

\author{Georg Jacob}
\email{georg.jacob@uni-mainz.de}
\affiliation{QUANTUM, Institut f\"ur Physik, Universit\"at Mainz, Staudingerweg 7, 55128 Mainz, Germany}

\author{Karin Groot-Berning}
\affiliation{QUANTUM, Institut f\"ur Physik, Universit\"at Mainz, Staudingerweg 7, 55128 Mainz, Germany}

\author{Sebastian Wolf}
\affiliation{QUANTUM, Institut f\"ur Physik, Universit\"at Mainz, Staudingerweg 7, 55128 Mainz, Germany}

\author{Stefan Ulm}
\affiliation{QUANTUM, Institut f\"ur Physik, Universit\"at Mainz, Staudingerweg 7, 55128 Mainz, Germany}

\author{Luc Couturier}
\altaffiliation{Present address: National Laboratory for Physical Sciences at Microscale and Department of Modern Physics, University of Science and Technology of China, 201315 Shanghai, China}

\author{Samuel T. Dawkins}
\affiliation{QUANTUM, Institut f\"ur Physik, Universit\"at Mainz, Staudingerweg 7, 55128 Mainz, Germany}

\author{Ulrich G. Poschinger}
\affiliation{QUANTUM, Institut f\"ur Physik, Universit\"at Mainz, Staudingerweg 7, 55128 Mainz, Germany}

\author{Ferdinand Schmidt-Kaler}
\affiliation{QUANTUM, Institut f\"ur Physik, Universit\"at Mainz, Staudingerweg 7, 55128 Mainz, Germany}

\author{Kilian Singer}
\affiliation{
Institut f\"ur Physik, Universit\"at Kassel, Heinrich-Plett-Stra{\ss}e 40, 34132 Kassel, Germany}

\date{\today}

\begin{abstract}
We realize a single particle microscope by using deterministically extracted laser cooled $^{40}$Ca$^+$ ions from a Paul trap as probe particles for transmission imaging. We demonstrate focusing of the ions with a resolution of 5.8$\;\pm\;$1.0$\,$nm and a minimum two-sample deviation of the beam position of 1.5$\,$nm in the focal plane. The deterministic source, even when used in combination with an imperfect detector, gives rise to much higher signal to noise ratios as compared with conventional Poissonian sources. Gating of the detector signal by the extraction event suppresses dark counts by 6 orders of magnitude. We implement a Bayes experimental design approach to microscopy in order to maximize the gain in spatial information. We demonstrate this method by determining the position of a 1$\,\mu$m circular hole structure to an accuracy of 2.7$\,$nm using only 579 probe particles.
\end{abstract}


\maketitle

\section{Introduction}

The advancement of electron microscopes~\cite{knoll1932elektronenmikroskop,muller1936versuche} and subsequently ion microscopes~\cite{muller1956FieldIonization,levisetti1975scanningIonMicroscope,orloff1975studyFieldIonization} has often been driven by new or improved types of sources. Together with improvements to imaging optics, better sources have helped to push the resolution of imaging far below the diffraction limit of visible light~\cite{ward2006helium,Erni2009AtomicResolution}, enabling substantial progress across various scientific~\cite{kruger2000helmut} and industrial~\cite{Bassim2014RecentAdvances} fields.

More recently, techniques pioneered in cold atoms have been employed to improve the sources in terms of phase space occupation, temporal control and offering deterministic emission properties. For example, laser cooled caesium ions extracted from a cold magneto-optical trap have been utilized to achieve high brightness and a nanometer spot size~\cite{knuffman2013cold}. Other approaches use Rydberg excitation to spatially shape and extract ultracold electron bunches from a magneto-optical trap with picosecond resolution~\cite{scholten2011arbitrarily}. A single-atom source has been realized, harnessing the two-body losses in a Bose-Einstein condensate~\cite{Manning2014Single-AtomSource}. Proposals for implementing single ion sources include using Rydberg atoms and the dipole blockade mechanism~\cite{ates2013fast,urban2009observation,miroshnychenko2009observation}, or an atomic ensemble within an optical lattice during the transition to the single occupancy Mott insulator~\cite{greiner2002quantum}. In addition an ultrafast electron source has been realized by extraction of electrons from a tungsten tip via femtosecond laser pulses\cite{Hommelhoff2006UltrafastElectronPulses,Hommelhoff2009ExtremeLocalization}. In this paper, we implement a single-ion source by extracting laser-cooled $^{40}$Ca$^+$ ions from a linear segmented Paul trap~\cite{schnitzler2009deterministic,izawa2010controlled} and use it to realize a novel type of microscopy based on deterministic probing.

In conventional microscopy, the signal-to-noise ratio\,(SNR) can typically be improved by increasing the exposure time or the flux. This is a direct consequence of the Poissonian statistics of the sources in use. However, a high particle emission can be detrimental in some applications where, for example, high irradiation causes charging~\cite{YoungMin2010Practical}, contamination or even damage~\cite{Prawer1995IonBeam} to samples. 
The approach presented here, addresses this problem in a fundamentally different way, namely by probing with a deterministic source. In principle, such a source could give rise to noiseless imaging, requiring an exposure of only a single particle to probe for transmission. However, in combination with a detector of finite quantum efficiency, the signal statistics become binomial. This still leads to inherently more information per particle and thus higher SNR than would be possible with Poissonian statistics. In addition, the source allows for gating the detection by the extraction event, yielding a suppression of the detector dark counts by six orders of magnitude. Finally, with a deterministic source, the Bayes experimental design method can be used to maximize the spatial information gained when imaging transmissive structures with a parametrizable transmission function.

A description of the source, as well as the entire microscope, is given in section~\ref{sec:Single ion microscope setup}, followed by a characterization in terms of achievable resolution and beam pointing stability in section~\ref{sec:Beam waist and beam stability of the single ion microscope}. In section~\ref{sec:Imaging with single ions} deterministic single ion microscopy is demonstrated. We compare images, which are acquired using deterministic and Poissonian sources and discuss the resulting SNR with respect to the total exposure. An information-driven approach to imaging based on the Bayes experimental design method is presented in section~\ref{sec:Bayes experimental Design}.

\section{Single ion microscope setup\label{sec:Single ion microscope setup}}

\begin{figure}
\begin{center}
\includegraphics[width=0.9\columnwidth]{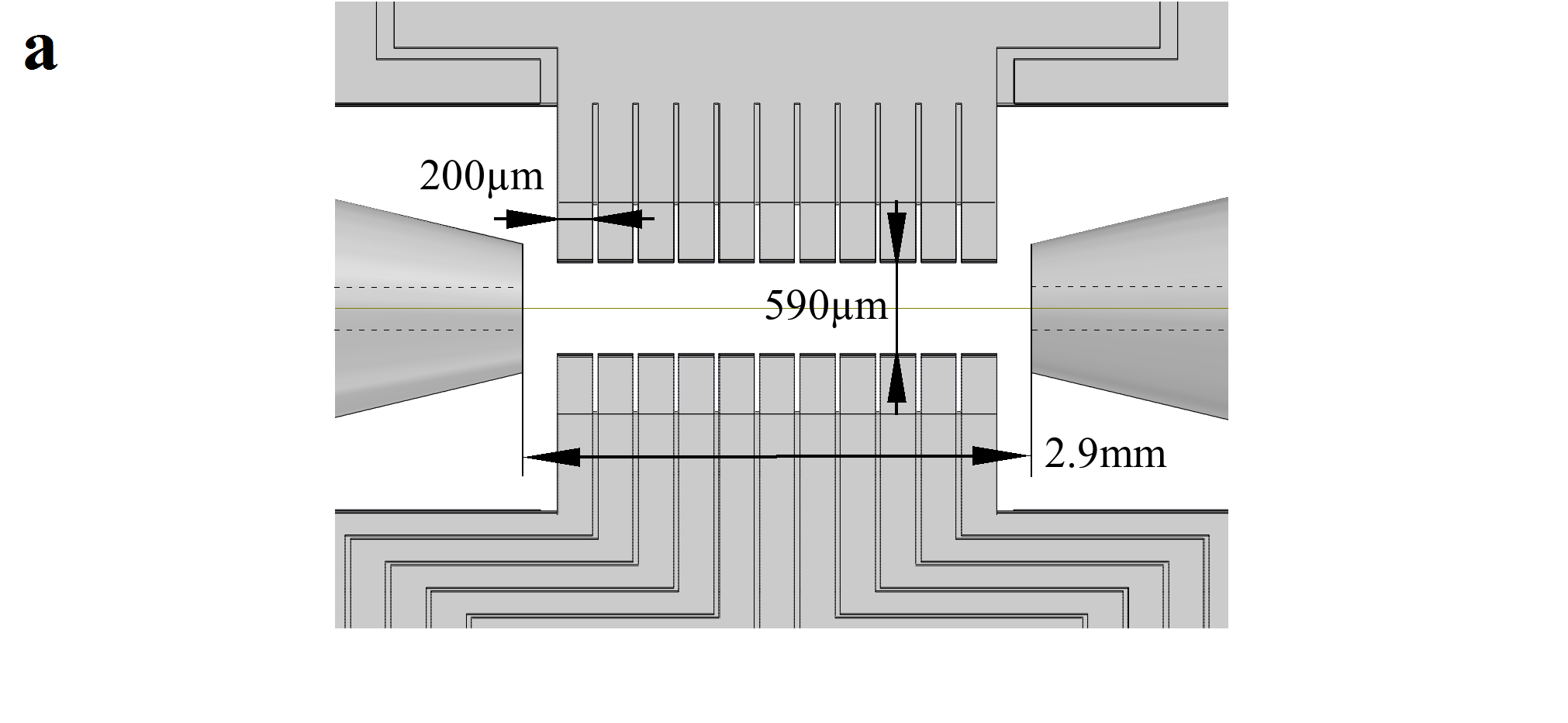}
\includegraphics[width=0.9\columnwidth]{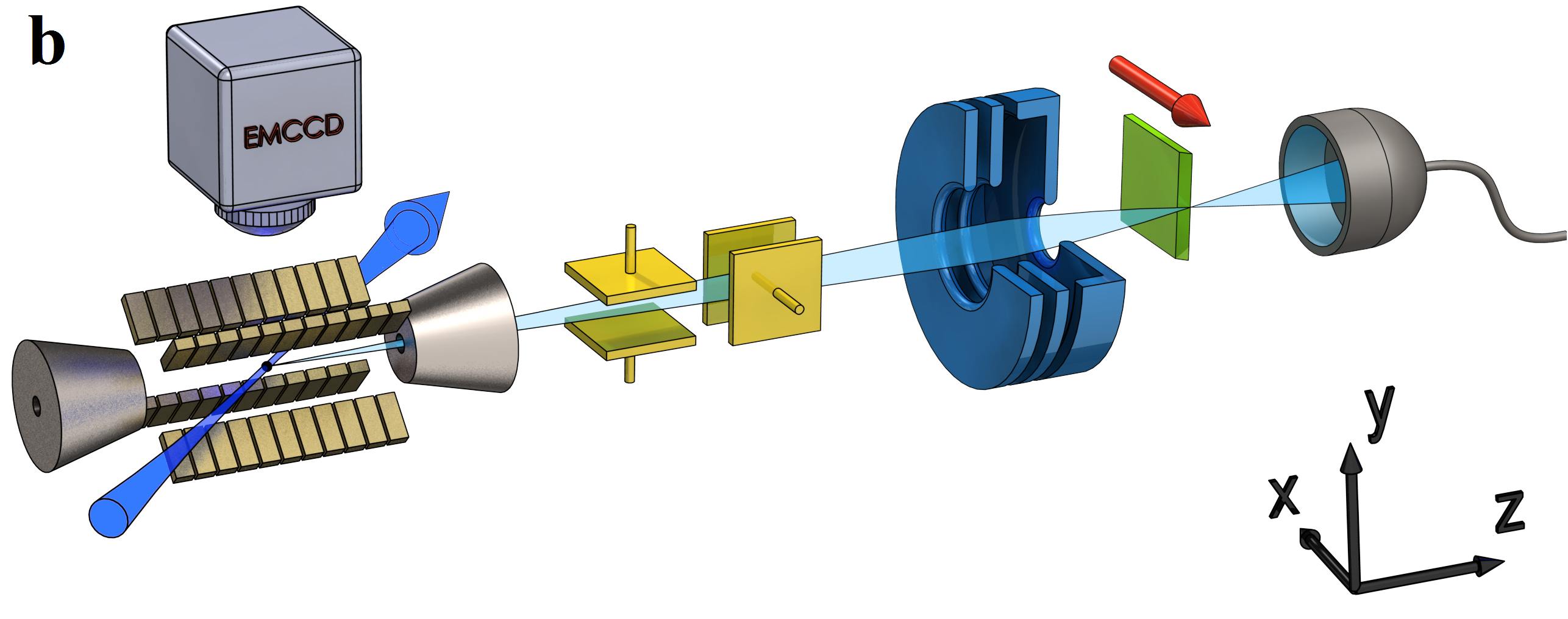}
\caption{a) True to scale image of the trap geometry indicating the important dimensions, as seen from the CCD-camera. b) Sketch of the single-ion microscope. The ion trap consists of segmented electrodes and end-caps. Laser-cooling (blue arrow) and imaging CCD system (from top) are shown. Deflection electrodes (yellow) and einzel-lens (blue) are placed along the ion extraction pathway (light-blue). In the focal plane a profiling edge or alternatively a transmissive mask (green) is placed on a three-axis nano-translation stage (not shown). Single ion counting is performed with a secondary electron multiplier device down stream.\label{fig:trap}}
\end{center}
\end{figure}

The experimental setup is based on a Paul trap which is constructed from four micro-fabricated alumina chips arranged in an X-shaped configuration and two metal end-caps (see Fig.~\ref{fig:trap}a). Each chip comprises 11 electrodes for shaping the confining electrostatic potential along the axial direction. We operate the device at trap frequencies  $\omega/(2\pi)$ of $0.58\,$MHz~to~$0.85\,$MHz and $1.4$\,MHz~to~$3.3\,$MHz for the axial and radial modes of vibration, respectively.
 
The deterministic source is implemented by a fully automated procedure:
Initially a random number of calcium ions are loaded by photo-ionization and laser-cooled on the S$_{1/2}$ to P$_{1/2}$ dipole transition near 397$\,$nm. The number of ions is counted by imaging the ion fluorescence on a CCD-camera and then reduced to the desired number by lowering the axial trapping potential with a predefined voltage sequence, which is applied to the trap segments (Fig.\ref{fig:trap}a). The cold ions are extracted along the axial direction of the trap by application of an acceleration voltage of up to -6$\,$kV to one of the pierced end-caps. For the experiments reported here, an extraction voltage of -2.5$\,$kV is used. Fast high voltage solid-state switches allow for a jitter of less than 1$\,$ns. The extraction time is synchronized to the phase of the radio-frequency trap-drive ($\Omega/(2\pi)=23\,$MHz) with adjustable delay. With this method we attain rates for loading and extraction of single ions of up to 3$\,$s$^{-1}$, corresponding to an average flux of about 0.5$\,$attoampere. Ions leave the trap passing through the 200$\,\mu$m diameter hole in the end-cap and are detected by a secondary electron multiplier with a quantum efficiency of about $96\,$\%. We measure a time-of-flight signal with a half-width half-maximum spread of $\Delta t=270\,$ps. This corresponds to a velocity spread of $\Delta v=8\,$m/s at a typical average speed of $10^5\,$m/s.

In order to align and scan the beam, two pairs of deflection electrodes are placed along the extraction path at a distance of 46$\,$mm and 67$\,$mm from the center of the trap respectively (see Fig.~\ref{fig:trap}b). For focusing of the beam, an electrostatic einzel-lens is placed 332$\,$mm from the center of the trap. It consists of three concentric ring shaped electrodes with an open aperture of 4$\,$mm. The geometry parameters are optimized by electrostatic simulations~\cite{Singer:2010} to minimize spherical aberration. Chromatic aberration is strongly suppressed due to the narrow velocity distribution of the ions. Image information is generated by recording transmission events for a well defined number of extractions while scanning the position of an object in the focal plane, using a three-axis translation stage.

\begin{figure}[h]
\begin{center}
\includegraphics[width=0.9\columnwidth]{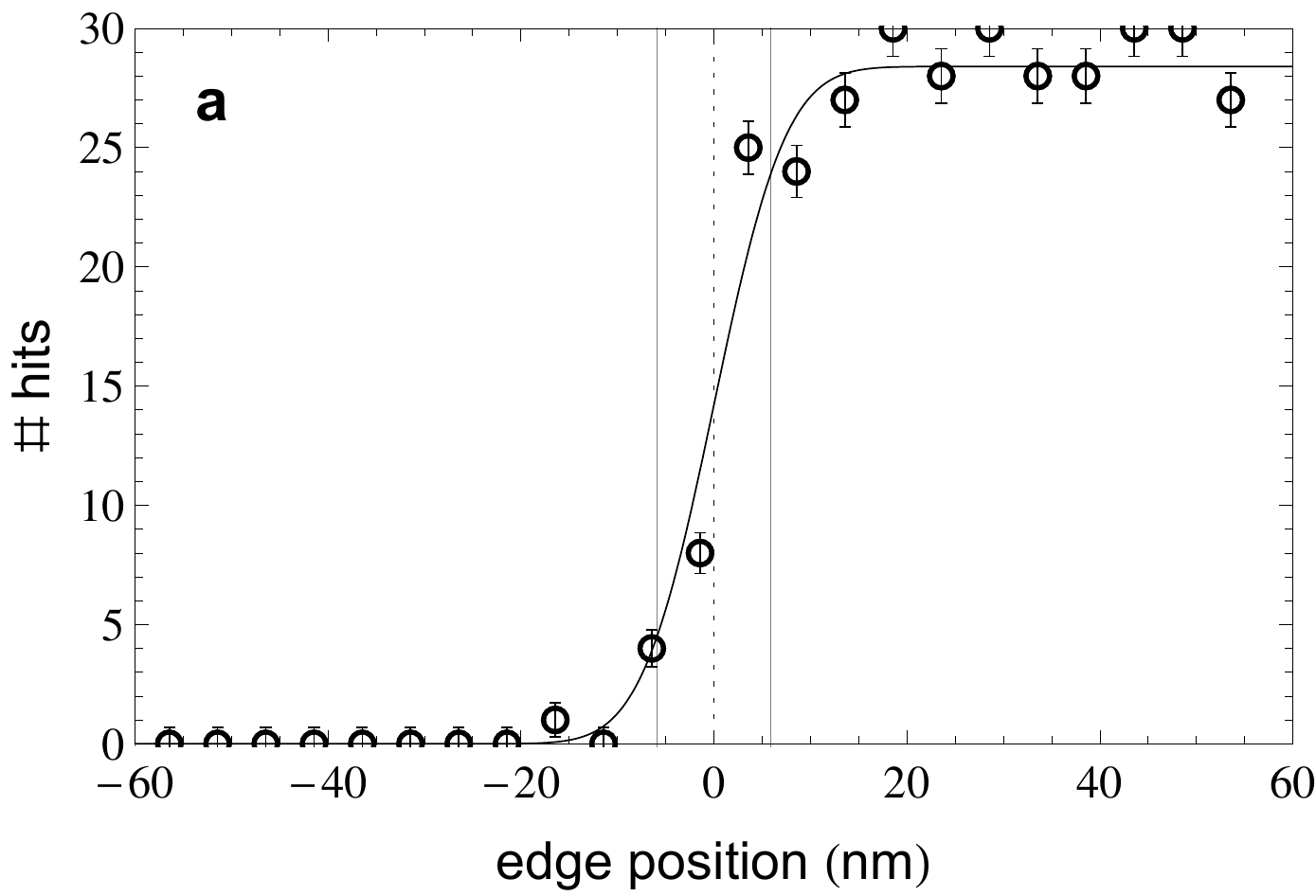}
\hfill
\includegraphics[width=0.9\columnwidth]{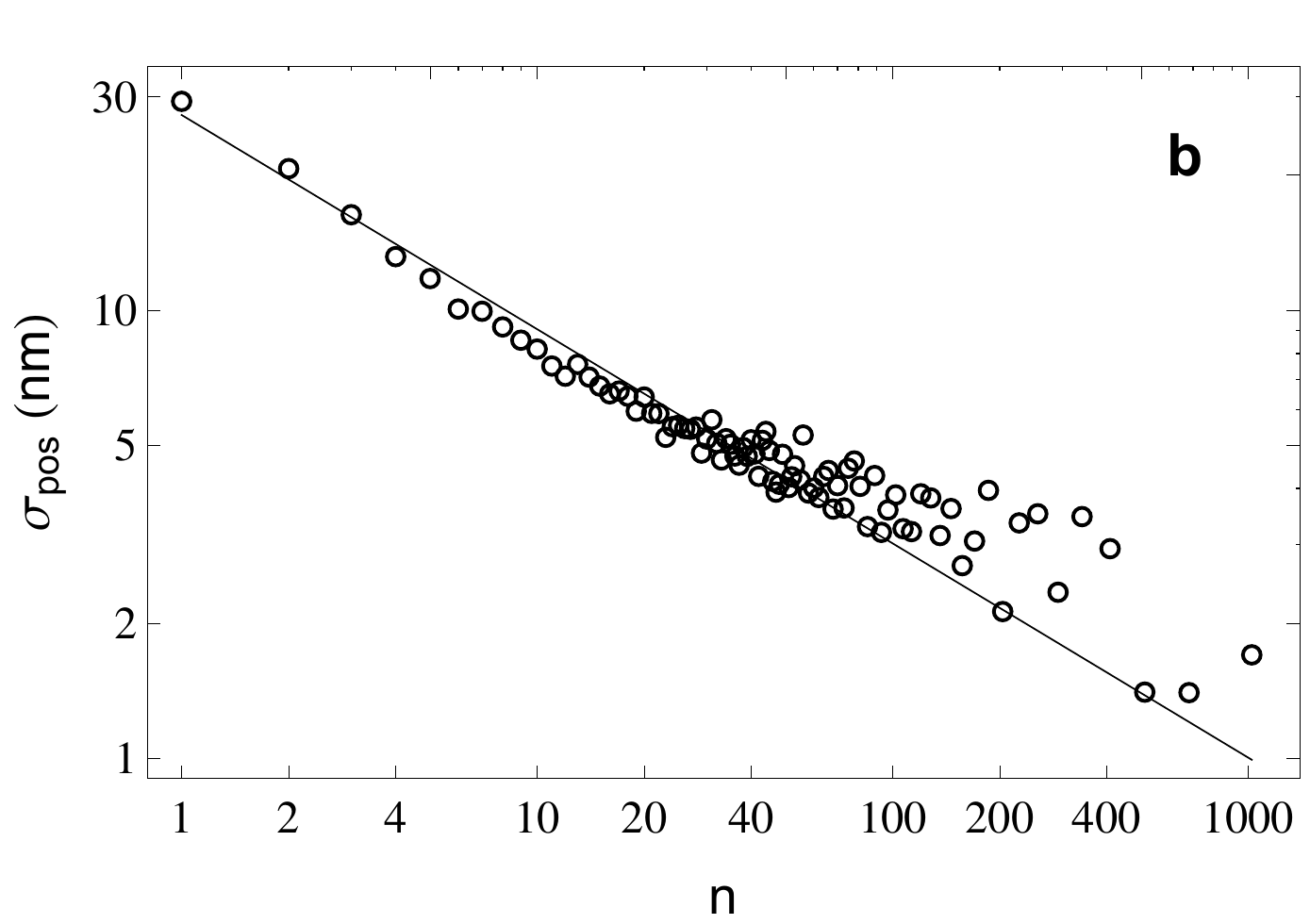}
\end{center}
\caption{
a) Number of detected ions out of 30 single ion extractions at the corresponding profiling edge position (circles). Errors are determined from binomial counting statistics. Dashed line shows center position and gray lines show 1-sigma radius of the beam waist. Fit with a Gaussian error-function $p(x)=\frac a2\left[1+\textrm{erf}\left(\frac{x-x_0}{\sigma\sqrt 2}\right)\right]$ to the data yields $a=\;$28.1$\;\pm\;$0.5 and $\sigma=\;$5.8$\;\pm\;$1.0$\,$nm.
\label{fig:focus} b) Log-log plot of the two-sample deviation of the beam position. The fit (black line) to the beam position deviations (circles) reveals a slope of -0.48$\;\pm\;$0.01. Integration time for $n$=10 is about 5 minutes, for $n$=100 it is about one hour. 
\label{fig:two-sample}}
\end{figure}

\section{Beam waist and beam stability of the single ion microscope\label{sec:Beam waist and beam stability of the single ion microscope}}

For an accurate determination of the spatial resolution of the beam in the focal plane, a profiling edge is stepped into the beam and a fixed number of transmission measurements are made at each position (Fig.~\ref{fig:focus}a).
To obtain the beam parameters, the data is fitted with a Gaussian error function. Under optimal operating conditions this yields 5.8$\;\pm\;$1.0$\,$nm for the 1-sigma radius of the beam waist. 

In addition, long-term stability of the focus position and waist is an essential prerequisite for high resolution imaging. We have evaluated the two-sample deviation of the lateral position of the focus (see Fig.~\ref{fig:two-sample}b), as per the Allan-deviation technique used for evaluating the stability of atomic clocks. To this end, 2,048 profiling edge measurements (similar to those in Fig.~\ref{fig:focus}a) were carried out repetitively. Every measurement comprises 26 contiguous profiling edge positions separated by 10$\,$nm, each probed with a single ion. In total the data set thus contains 53,248 extraction events within an acquisition time of 18 hours. The two-sample variance (see Fig.~\ref{fig:two-sample}b) is given by

\begin{equation*}
\sigma^2_{\text{pos}}(n) = \frac{1}{2(N-1)}\sum_{i=0}^{N-1}\left( \bar{x}_{i+1}(n)-\bar{x}_i(n)\right)^2,
\end{equation*}

where $\bar{x}_i(n)$ is the beam position for the $i$-th segment, derived from fitting to the aggregate count data from $n$ consecutive profiling edge measurements.

If the measurements are dominated by statistical fluctuations rather than beam pointing drifts, the two-sample deviation scales as $1/\sqrt{n}$. We indeed observe a scaling exponent of -0.48(1), which demonstrates the long-term stability of the ion beam over the entire period of about nine hours.
Under stable thermal and constant trapping conditions~\footnote{We continuously heat the calcium oven and avoid the use of a titanium sublimation pump. The radio frequency-drive and all lasers are continuously switched on.}, the minimal two-sample deviation of the beam position yields a long-term beam-pointing stability of 1.5$\,$nm (Fig.~\ref{fig:two-sample}b).

\onecolumngrid
\begin{center}
\begin{figure}[h]

\includegraphics[width=0.25\columnwidth]{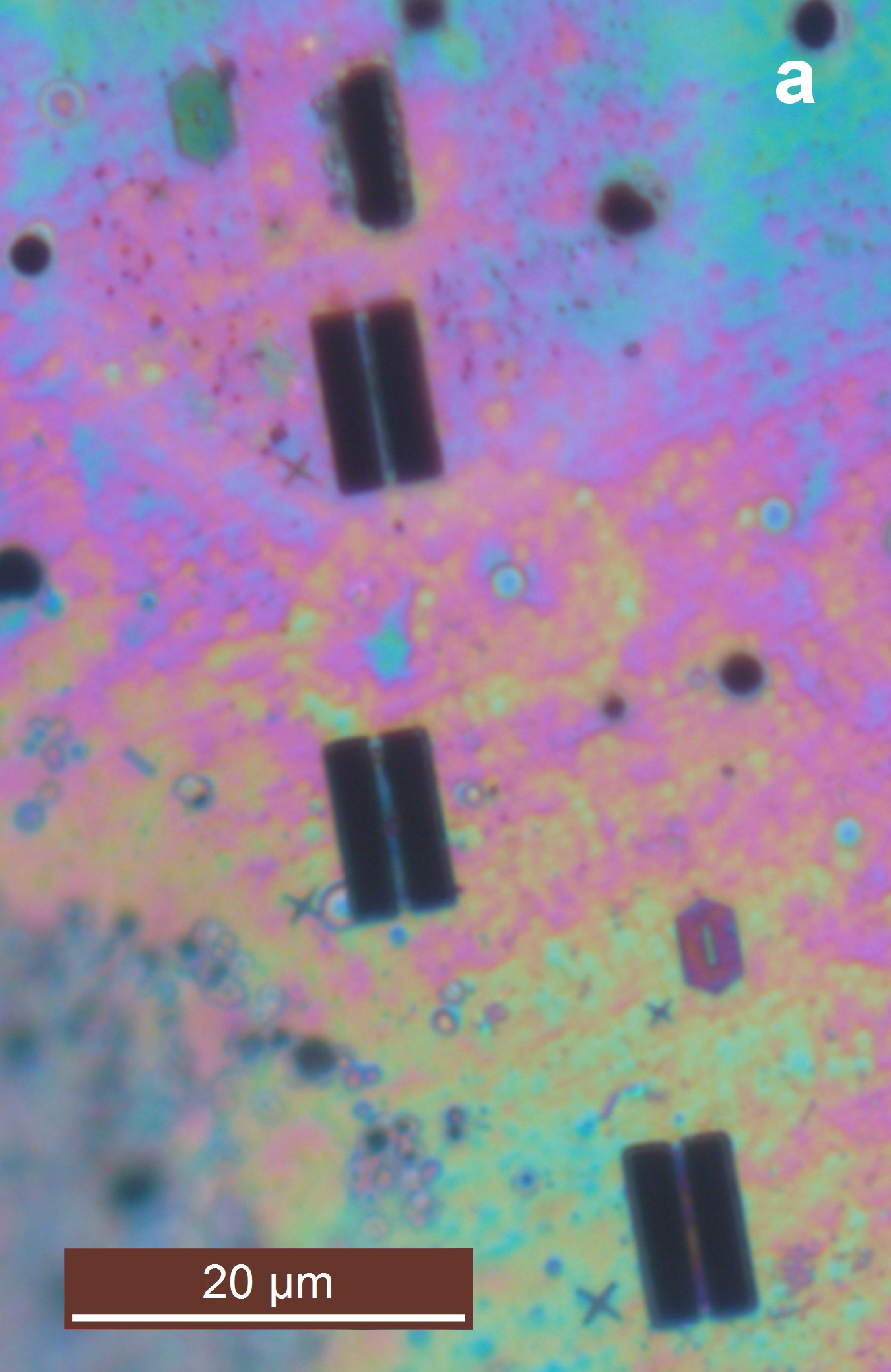}
\hfill
\includegraphics[width=0.215\columnwidth]{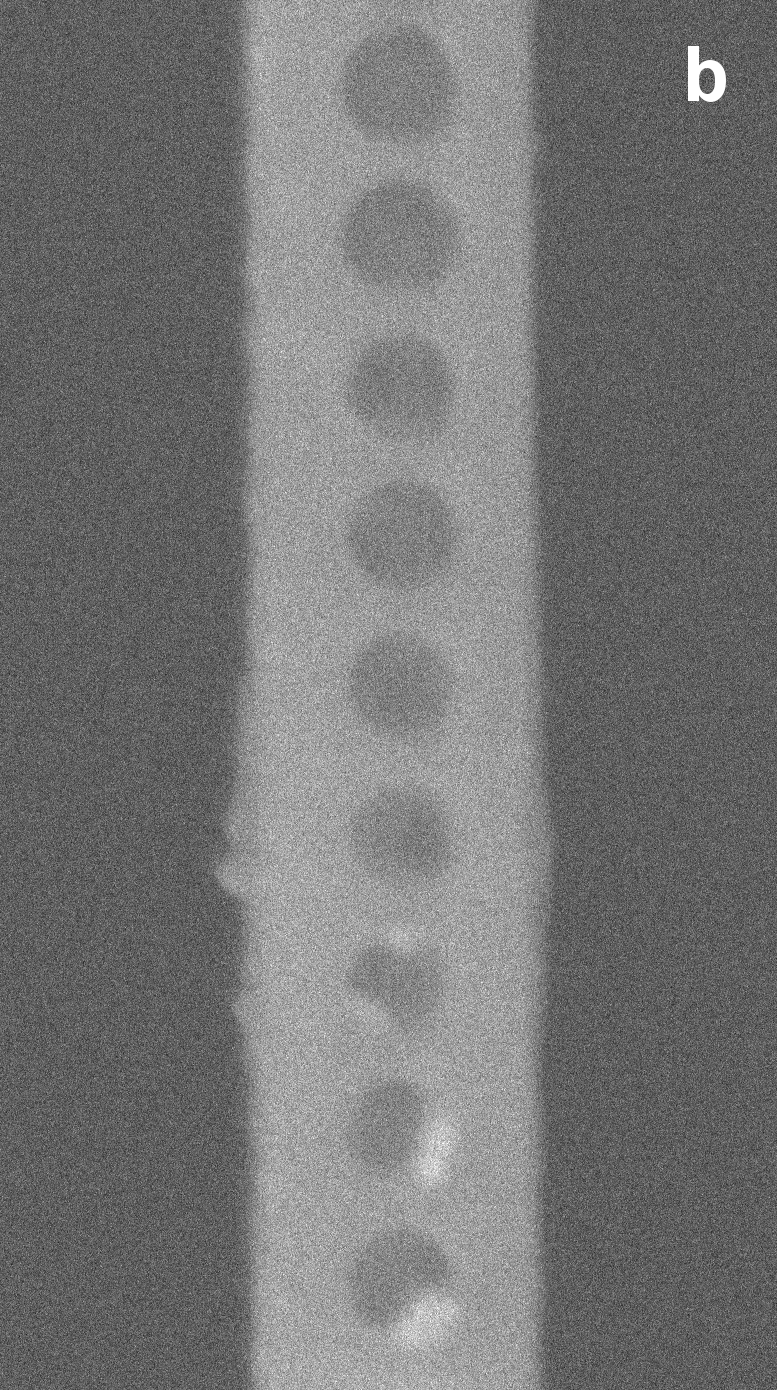}
\hfill
\includegraphics[width=0.207\columnwidth]{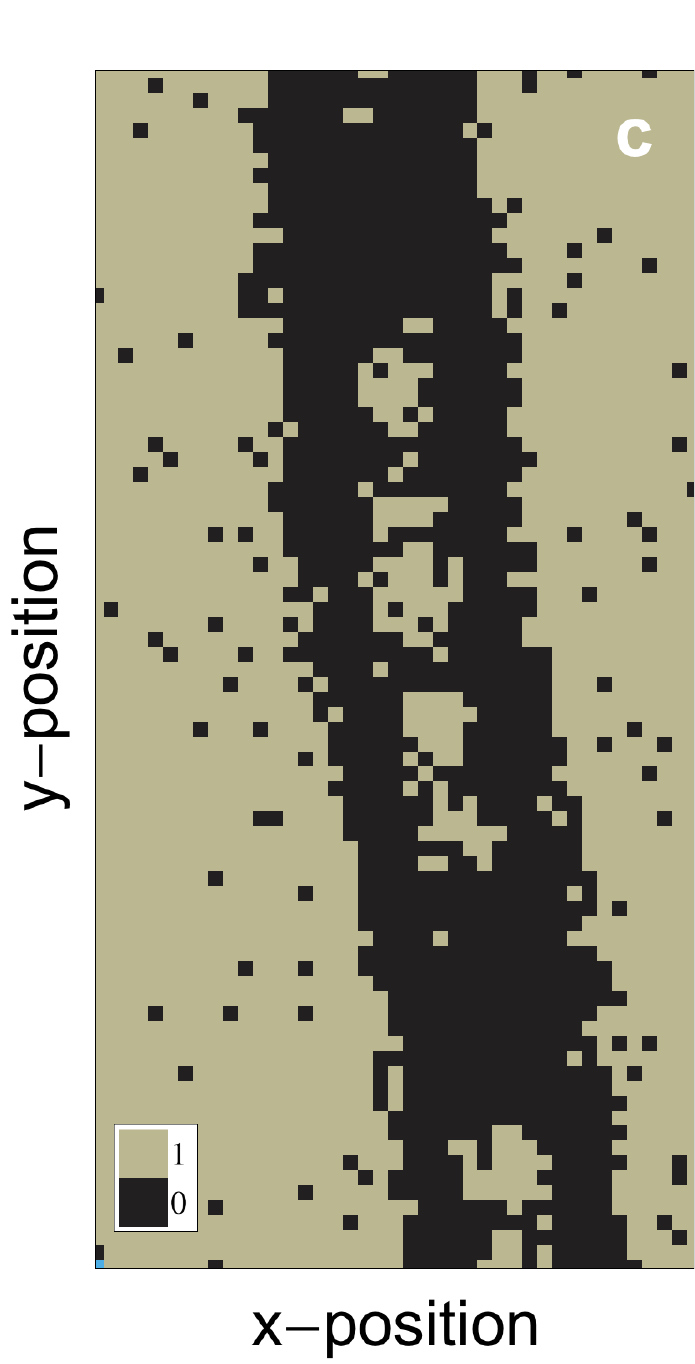}
\hfill
\includegraphics[width=0.207\columnwidth]{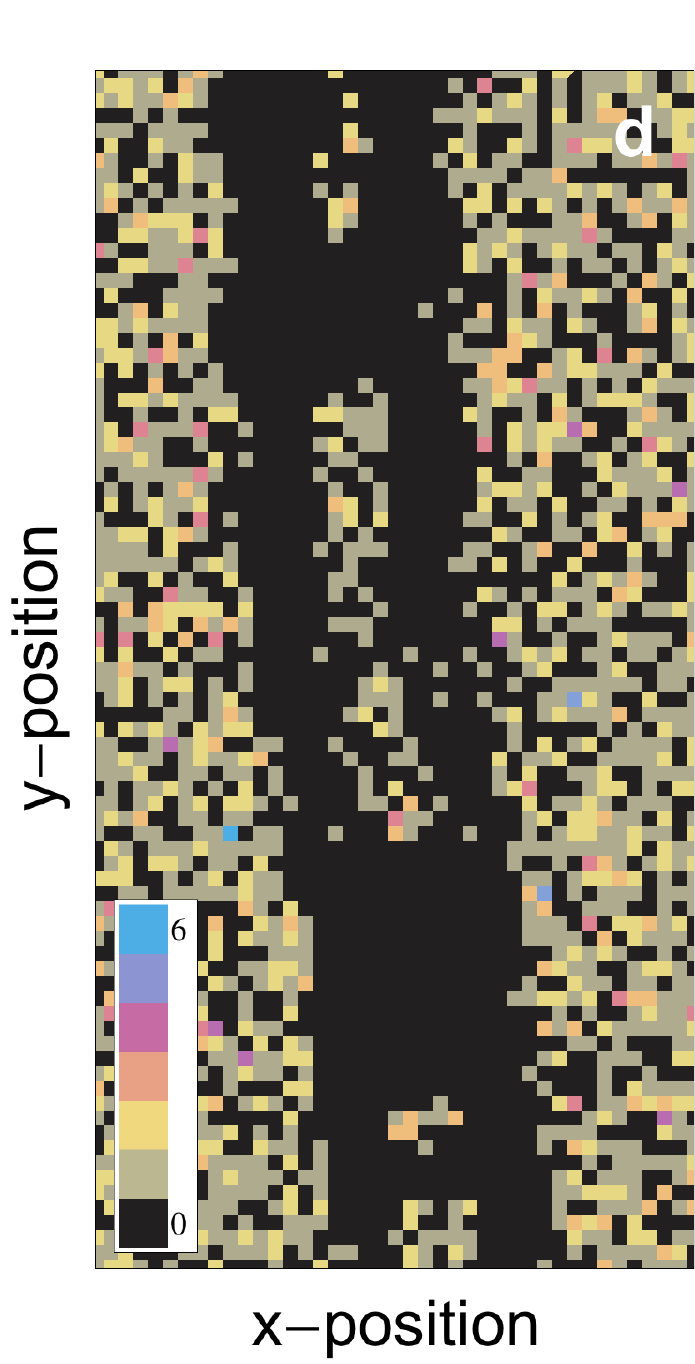}

\caption{Imaging a transmissive diamond sample: a) A large-area view of the transmissive structures (black) in the diamond sample~\cite{riedrich2012one}, imaged with a conventional optical microscope. The structure shown in (b), (c) and (d) is located in the lower right corner. b) SEM image of the waveguide-cavity structure. Holes have a diameter of about 150$\,$nm. c) The cavity structure is scanned using one ion at each lateral position, with a resolution of (25x25)$\,$nm$^2$ per pixel. The entire information in the picture is based on 2659 transmission events out of 4141 extracted ions. d) The same structure as imaged using a source with emulated Poissonian behaviour. The lower SNR as compared to (c) is clearly visible. The missing holes compared to the image in (b) are attributed to blind holes. Here the image is based on 2420 transmission events out of 3694 extracted ions.
\label{fig:scanplot}}
\end{figure}
\end{center}
\twocolumngrid

\section{Imaging with single ions\label{sec:Imaging with single ions}}

We demonstrate the imaging of transmissive structures by scanning a photonic waveguide-cavity fabricated from diamond (see Fig.~\ref{fig:scanplot}). We received this sample (300$\,$nm thickness, fabricated with Ga$^{+}$ ion FIB) from the group of C. Becher~\cite{riedrich2012one}. Fig.~\ref{fig:scanplot}a) shows a large-area view of the diamond substrate with these structures. A close-up SEM image of one of the structures is shown in Fig.~\ref{fig:scanplot}b). 

For acquisition of Fig.~\ref{fig:scanplot}c) each pixel is probed with exactly one ion. Gating the detector by the extraction event (gate-time typically less than 200$\,$ns) ensures that the dark count rate ($<\;$100 $s^{-1}$) does not affect the image contrast. Whereas the contrast in the transmissive areas is assumed to be limited solely by the detector efficiency, resulting in binomial counting statistics. The detector efficiency was measured to be (96$\pm$2)$\,$\%. Other influences such as background gas collisions are considered to be negligible. At the edges of the structure \textit{i.e.} between transmissive and non-transmissive parts, the contrast is dominated by the finite beam radius (compare Fig.~\ref{fig:focus}a). To contrast the imaging properties of the deterministic with a Poissonian source, Fig.~\ref{fig:scanplot}d) shows the same structure, but imaged with an experimentally emulated Poissonian source: Prior to the probing of each pixel, the number of ions to be extracted is obtained using a random number generator with a Poisson-distributed output, where the mean value is set to one.

\begin{center}
\begin{figure}
\includegraphics[width=\columnwidth]{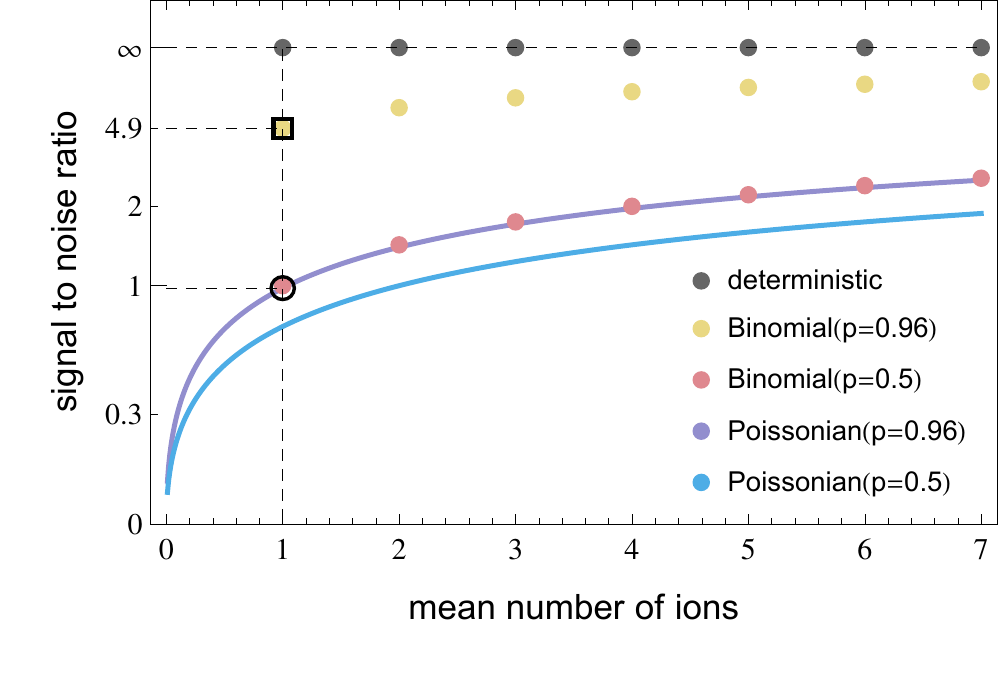}
\caption{Calculated SNR of a binomial and a Poissonian source plotted as a function of the mean number of extracted ions for different detector efficiencies. The scale on the y-axis is compactified by the function $f/(f+1)$. Note that the SNR for binomial statistics with $p=0.5$ is identical to the Poissonian SNR with $p=1$. The square and the circle depict the operating points of the detector-efficiency-limited deterministic source (as in Fig.~\ref{fig:scanplot}c) and the emulated source (as in Fig.~\ref{fig:scanplot}d) respectively. In both cases the mean number of extracted ions per pixel is one and the detector efficiency is $0.96$. 
This results in an SNR of $4.90$ for the deterministic source and an SNR of $0.96$ for the emulated Poissonian source. 
\label{fig:scanplotDiscuss}}
\end{figure}
\end{center}

A quantitative comparison of the two types of sources is presented in Fig.~\ref{fig:scanplotDiscuss}. The SNR calculated as a function of the mean number of extracted ions is compared for different detection probabilities $p$. Note that the dark count noise is not taken into account in this comparison. This is justified by the aforementioned circumstance that the detector signal of both sources - the deterministic and the emulated Poissonian - is gated. For the plot the definition SNR~$=\mu/\sigma$ is used, where $\mu$ is the mean value and $\sigma$ the standard deviation of the corresponding probability mass function.

\onecolumngrid
\begin{center}
\begin{figure}[h]
\includegraphics[width=0.32\columnwidth]{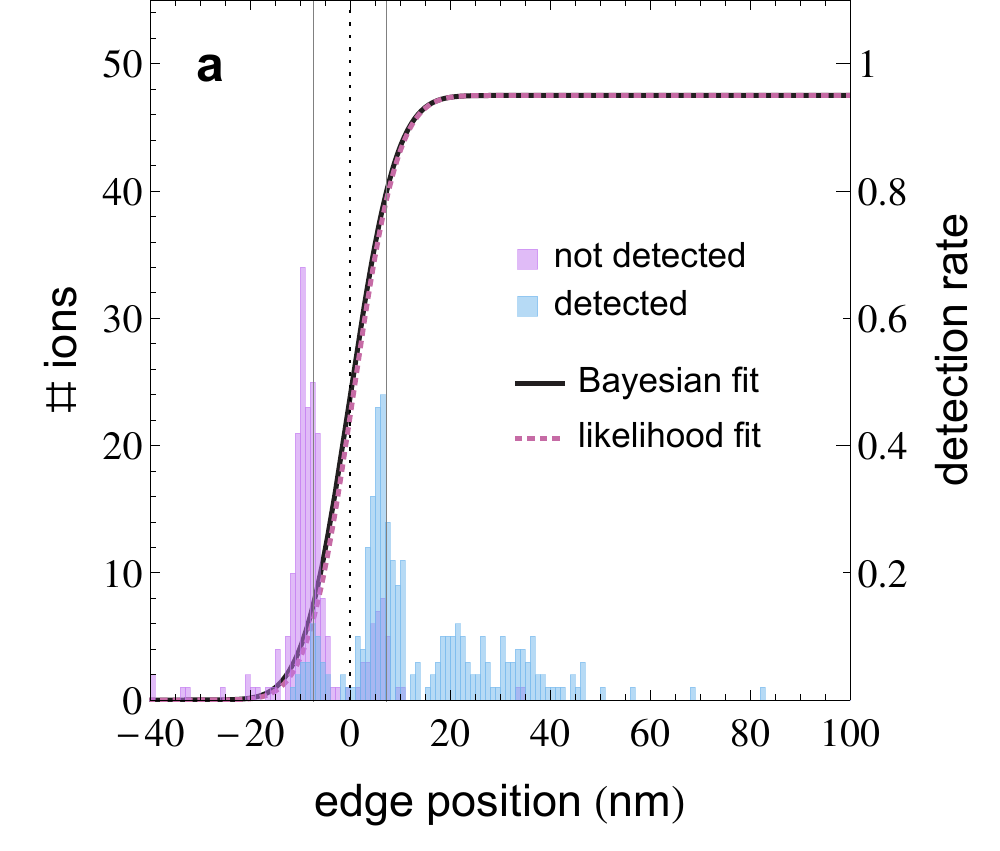}
\hfill
\includegraphics[width=0.32\columnwidth]{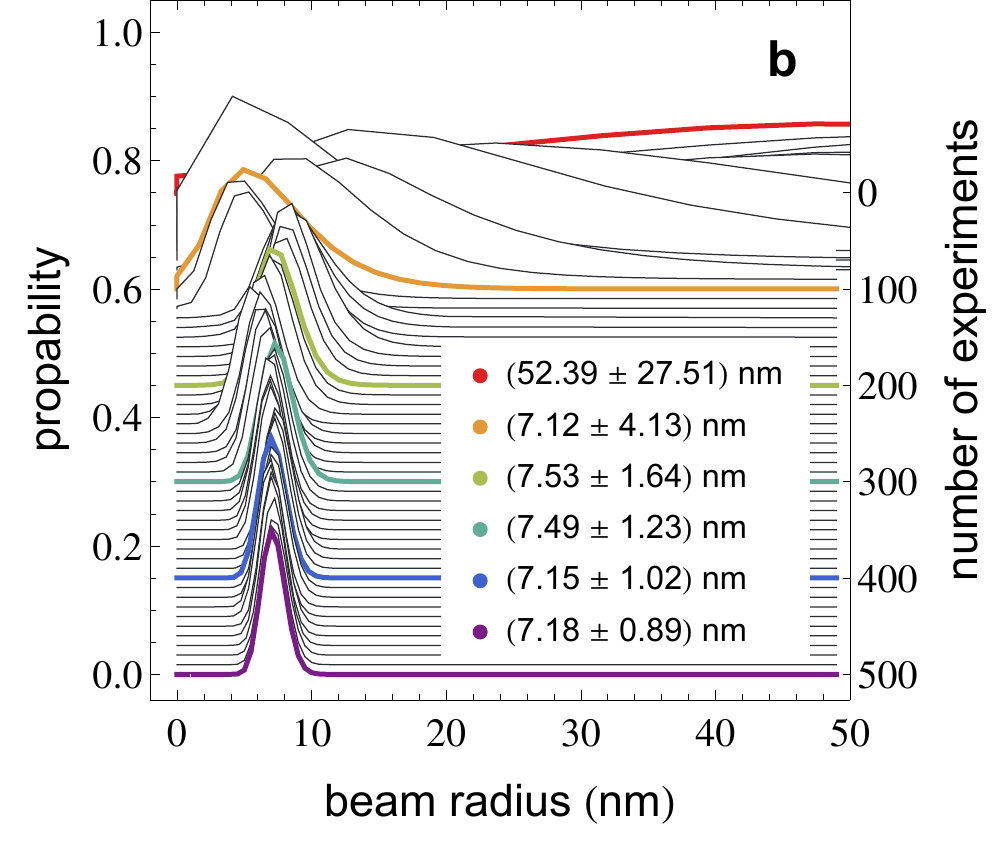}
\hfill
\includegraphics[width=0.32\columnwidth]{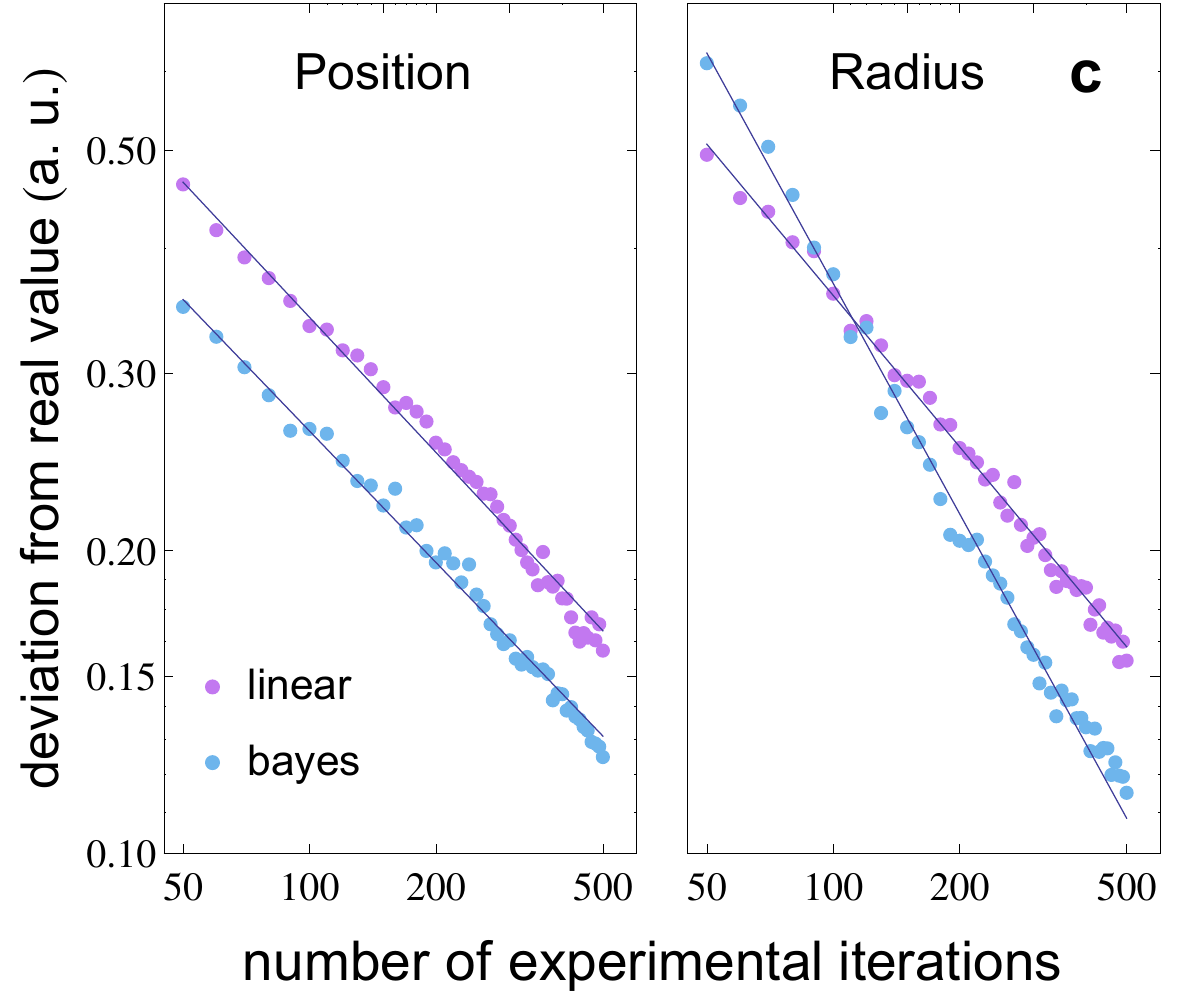}
\caption{Profiling edge measurement using the Bayes experimental design method: a) The histogram of optimal blade positions as calculated by the algorithm during the measurement. This data from 500 events is split into cases where the ion was detected~(blue) and cases where it was not~(purple). The Bayes fit function $p(y=1|\theta,\xi)$ is shown according to the final parameter values~(black), radius~$\sigma=7.18\pm0.89\,$nm and detector efficiency~$a=0.95\pm0.02$, the zero of the x-axis is set to $x_0$. For comparison, the result of a maximum likelihood fit~(dashed, purple) to the entire data, with $\sigma=7.13\pm0.66\,$nm and $a=0.95\pm0.02$, $x_0=0.73\,$nm. b) Marginal PDF of the beam radius as a function of the number of experimental iterations in the Bayes method. Only every tenth iteration is shown. c) Comparison of the stepwise and the Bayesian method by numerical simulations. The plot depicts the average deviation of the simulation results from the real value as a function of the number of iterations with the respective method. For each data point the deviation was calculated from the results of 1000 independent simulation runs. A multiplicative speed up of a factor of $\approx$\,4/3 is found, in determining the beam-position and an exponential speed up from $n^{-0.5}$ to $n^{-0.76}$ in determining the radius, when comparing the Bayesian method with the stepwise method.
\label{fig:BayesEdge}}
\end{figure}
\end{center}
\twocolumngrid

\section{Bayes experimental Design\label{sec:Bayes experimental Design}}

To maximize the spatial information gain per probe event, we make use of the deterministic nature of our source by using the Bayesian experimental design method~\cite{lindley1956measure,guerlin2007progressive,pezze2007phase,brakhane2012bayesian}. Employing this technique, it is possible to measure parameters of one or two-dimensional structures with a parametrizable transmission function. We first introduce the method by means of the profiling edge measurement from section~\ref{sec:Beam waist and beam stability of the single ion microscope} and demonstrate how the radius as well as the position of the beam can be obtained more efficiently as compared to the stepwise profiling method. In a second example, an algorithm is presented which is able to find and determine the lateral position of a circular hole structure with optimal efficiency.

In the Bayesian approach to parameter estimation, the knowledge about the value of a parameter $\theta$, given by pre-existing information, is expressed by the \textit{prior} probability distribution function (PDF) $p(\theta)$. Information from the outcome $y$ of a new measurement is subsequently incorporated using the Bayes update rule, yielding a \textit{posterior} PDF:
\begin{equation}
p(\theta|y,\xi)=\frac{p(y|\theta,\xi)p(\theta)}{p(y|\xi)}.
\label{eq:Bayes update rule}
\end{equation}
Here, the right hand side is the product of the prior PDF and the statistical model of the measurement $p(y\vert\theta,\xi)$, which is the probability to observe a outcome $y$ given the parameter values $\theta$ and design parameters $\xi$. $\xi$ contains the free control parameters of the experiment. Normalization is provided by the marginal probability of observing $y$, $p(y|\xi)=\int p(y|\theta,\xi)p(\theta) d\theta$.

The Bayes experimental design method consists in maximizing the information gain per measurement by the appropriate choice of the design parameters. The information gain of a measurement with outcome $y$ and control parameters $\xi$ is expressed by the \textit{utility} $U(y,\xi)$, which is the difference in Shannon entropies of the posterior and prior PDFs:
\begin{equation*}
U(y,\xi)= \int \ln(p(\theta|y,\xi))p(\theta|y,\xi) d\theta-\int \ln(p(\theta))p(\theta) d\theta.
\end{equation*}
Averaging the utility over the measurement outcomes yields a quantity independent of the hitherto unknown observation:
\begin{equation}
U(\xi)=\sum_{y\in\lbrace 0,1 \rbrace} U(y,\xi)p(y|\xi),
\label{eq:utility}
\end{equation}
which can be optimized with respect to $\xi$. Carrying out the measurement with control parameters $\xi$ thus ensures optimal information gain. 

For the case of the profiling edge measurement, the design parameter is the profiling edge position, while the parameters to be determined are the beam position $x_0$, its radius $\sigma$ and the detector efficiency $a$, \textit{i.e.} $\theta=(x_0,\sigma,a)$. The outcome of the measurement is binary, $y=\{0,1\}$. The measurement is modelled as
\begin{equation*}
p(y|\theta,\xi) = \begin{cases} 
\frac{a}{2} \, \text{erfc}\left[ \frac{\xi - x_0}{\sigma \sqrt{2}} \right] & \quad \text{if } y=1\\
1-\frac{a}{2} \, \text{erfc}\left[ \frac{\xi - x_0}{\sigma \sqrt{2}} \right] & \quad \text{if } y=0\\
\end{cases},
\end{equation*}
which in this case is a convolution of the transmission function of the structure to be imaged and a Gaussian beam profile.

The experimental sequence is carried out as follows. The initial prior, a three dimensional joint PDF, for the parameters $x_0$, $\sigma$ and $a$, is chosen. Its marginals can be uniform or an educated guess \textit{e.g.} a Gaussian distribution. It is implemented numerically, being a three dimensional grid of equidistant, weighted and normalized sampling points. For updating the prior to allow for the utility calculation~(\ref{eq:utility}), the Bayes update~(\ref{eq:Bayes update rule}) is performed individually for each sampling point. The maximizing algorithm is realized by calculating the utility for an interval divided into equidistant profiling edge positions and recursively repeating this calculation for a smaller interval around the maximum. Five recursions were found to be sufficient to reach the required accuracy without incurring excessive computational expense. Here, the integrals are replaced by sums over all sampling points. Using the measurement outcome of the real experiment performed at the calculated optimal profiling edge position, the Bayesian update~(\ref{eq:Bayes update rule}) is applied to calculate the actual posterior PDF, which assumes the role of the prior PDF for the next iteration. The procedure is repeated until an accuracy goal is reached. 

Fig.~\ref{fig:BayesEdge}a) shows the result of a typical Bayes-optimized profiling edge measurement. The parameter values for Bayes fit function are derived by calculating the mean values of the marginal PDFs of the corresponding parameters. For comparison, a maximum likelihood fit is also shown, since the values determined by the Bayesian method are in principle not independent of the exact sequence. Fig.~\ref{fig:BayesEdge}b) shows how the marginalized PDF for the beam radius $\sigma$ changes while the iteration proceeds. To compare the stepwise method with the Bayesian method, we implement numerical simulations of both approaches and calculate the average deviation of the si\-mulation outcome from the real value, as a function of the number of iterations $n$. The result is depicted in Fig.~\ref{fig:BayesEdge}c).

\begin{center}
\begin{figure}[h]
\includegraphics[width=0.31\columnwidth]{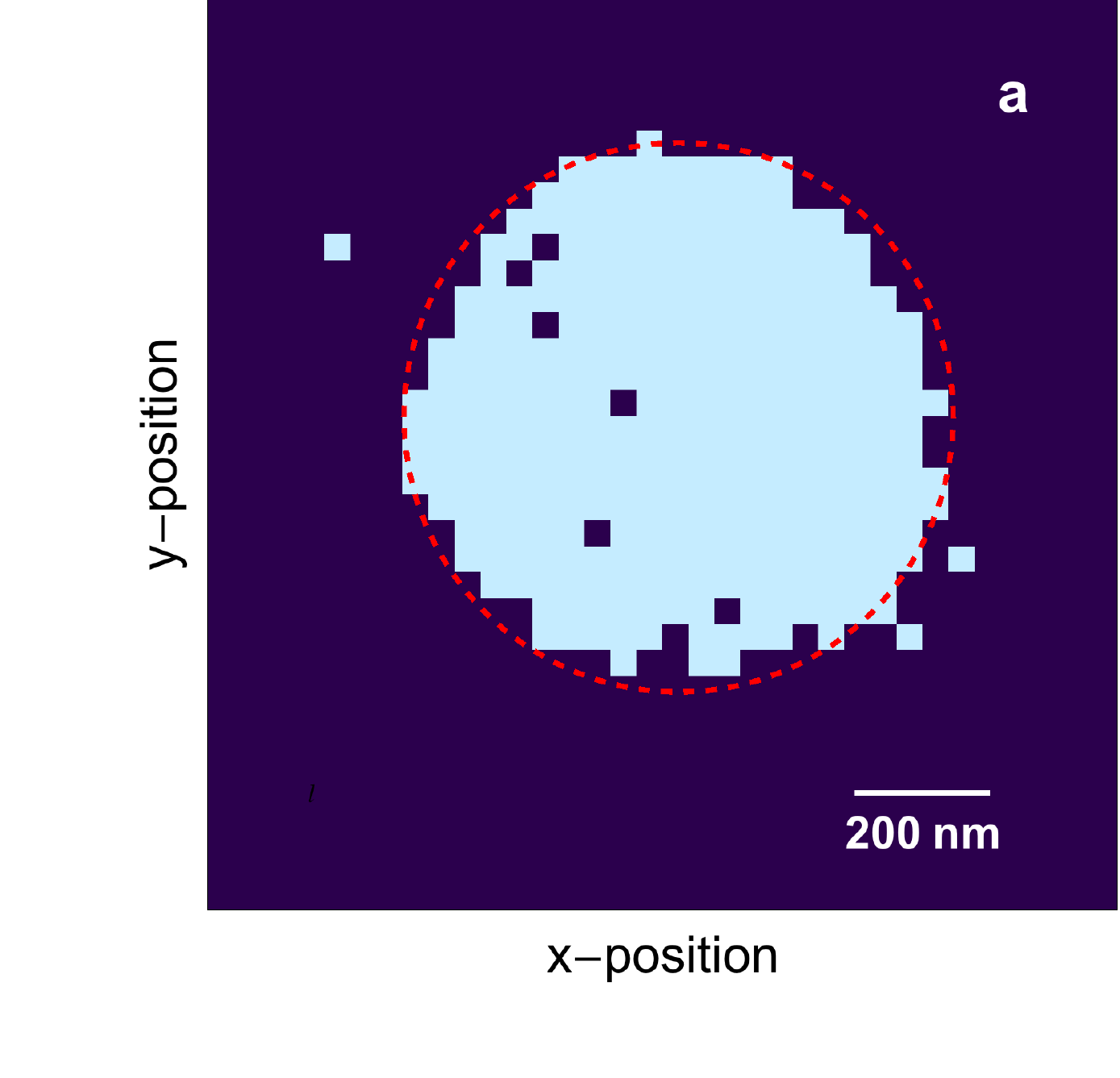}
\hfill
\includegraphics[width=0.31\columnwidth]{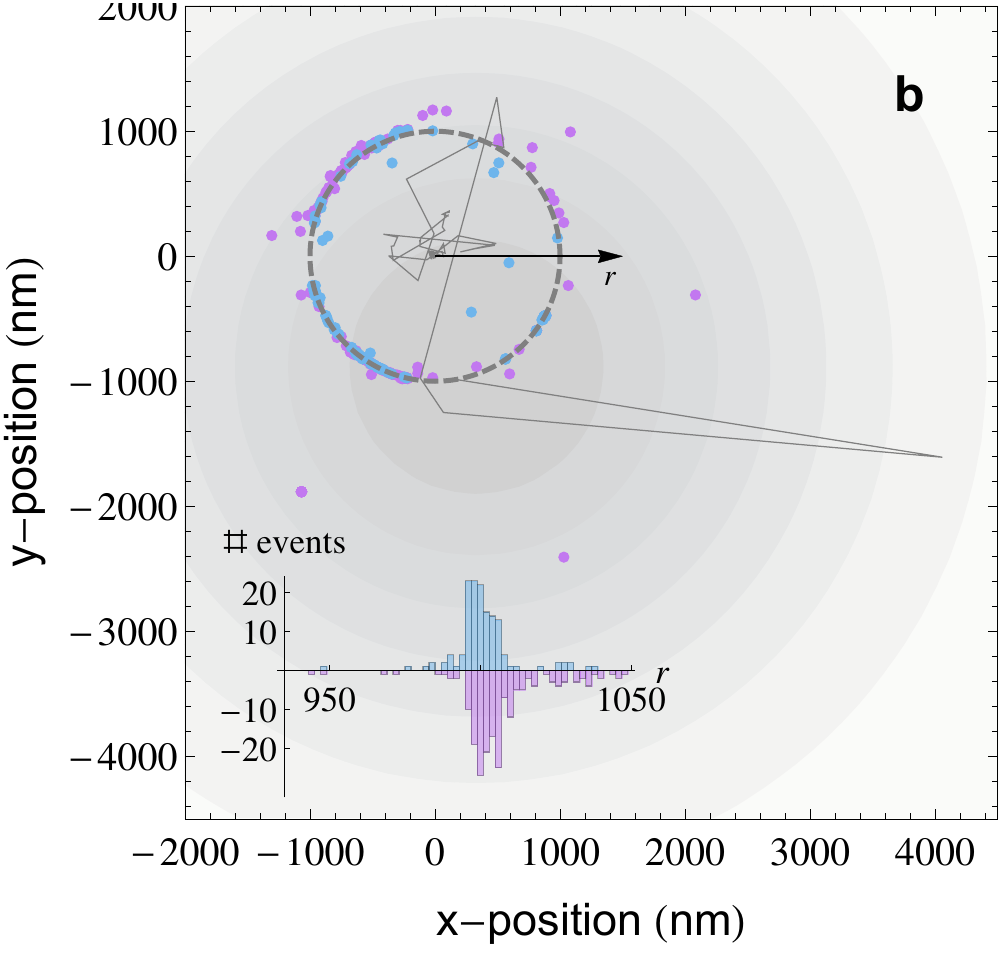}
\hfill
\includegraphics[width=0.34\columnwidth]{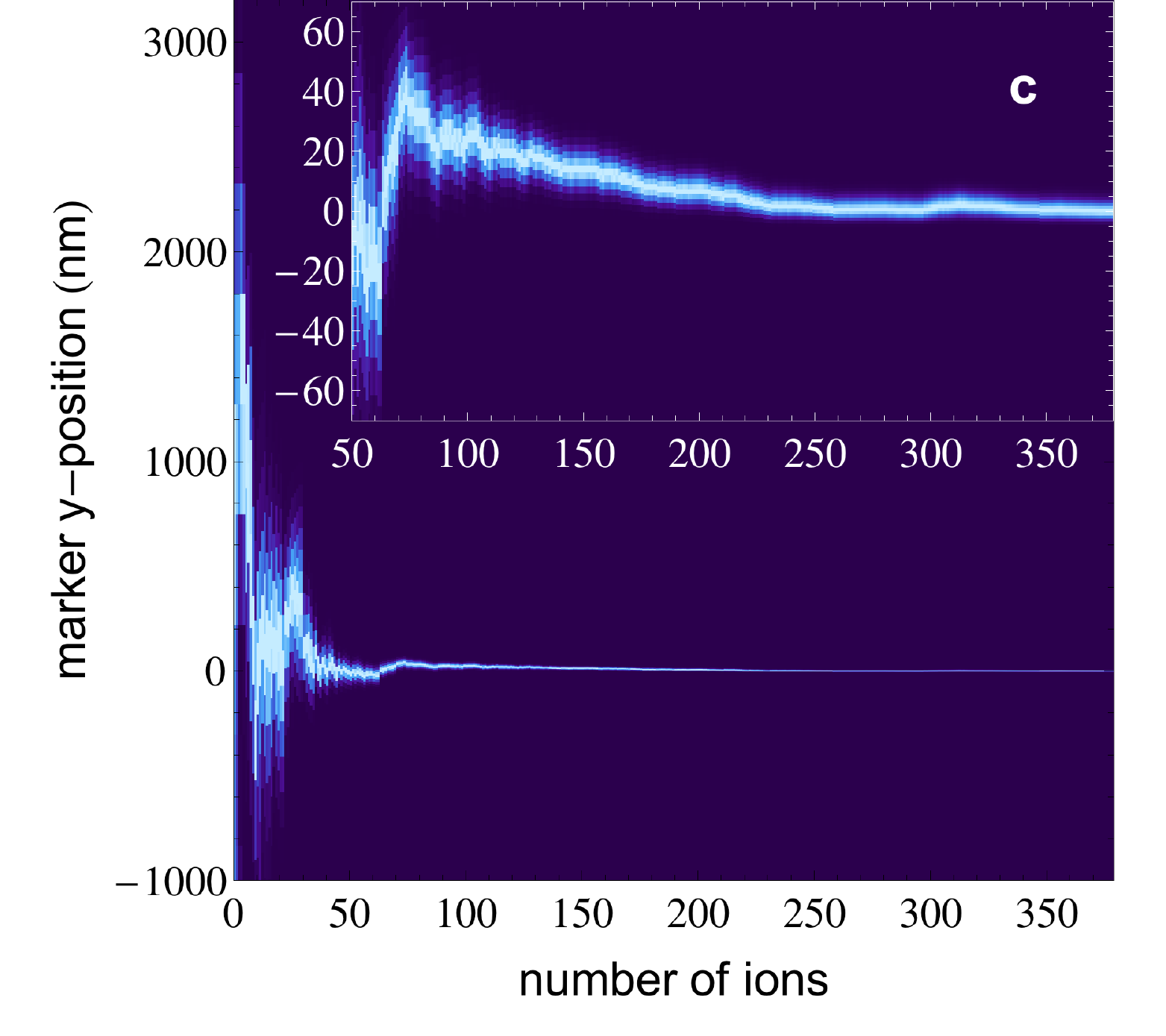}
\caption{Precise determination of alignment-hole parameters: a) A hole structure is scanned using one ion at each lateral position, with a resolution of (100x100)$\,$nm$^2$ per pixel. The red circle shows the result of a maximum likelihood fit to the data. b) The same structure is measured using the Bayes experimental design method.  For the plot, the $x$ and $y$ position of the hole were set to zero. The blue and red dots represent the positions where an ion was, or was not, detected, respectively. The final location and radius of the hole structure is depicted by the dashed circle. The inset shows a histogram of detected and not detected events dependent on $r$ the distance to the center of the structure. The initial guess of the Gaussian shaped PDF for the position is depicted as a gray shade in the background. The dark gray line follows the progression of its mean value \textit{i.e.} the assumed center position of this distribution as a function of the number of extracted ions. Within the first four iterations no ion is transmitted. The spatial information of these blocked particles shifts the assumed position, since it excludes that specific areas are transmissive. After the first ion is transmitted the assumed position makes a step towards its location. c) Plot of the marginal PDF for the $y$ coordinate of the circle.  With the first transmitted ion, the width of the PDF collapses, because the position of the structure is now known to within its assumed radius. The inset shows a zoom to the region around the final value.
\label{fig:BayesMarker}}
\end{figure}
\end{center}

We demonstrate the measurement of parameter values of two-dimensional transmissive structures with a parametrizable transmission function by means of a circular hole in a diamond sample (see Fig.~\ref{fig:BayesMarker}). This is also a practical example for sample alignment, since for many applications it is useful to know the exact lateral position of a sample with respect to the beam focus. For this purpose two perpendicular profiling edges as used in the previous example could equally be employed. However, for practical reasons, it might be more convenient to use a simple hole structure, which is in close proximity to a structure of interest, as a marker.
	
For comparison the hole structure  is first scanned with a linear sequence, using 1332~ions in total, where each lateral position is probed with one ion. A maximum likelihood fit to the data yields an accuracy of $\Delta x=47.1\,$nm and $\Delta y=22.6\,$nm for the position, where the radius was extracted to be $r=1057.0\pm31.8\,$nm. For applying the Bayesian method, the experiment is parametrized by the lateral position of the center of the circular hole, its radius as well as the 1-sigma beam radius and the detector efficiency. The radius of the beam and the detector efficiency were kept constant at 25$\,$nm and 95$\,$\% respectively. Both values were measured separately in advance. Using 379 ions in total, the position was determined with an accuracy of $\Delta x=2.7\,$nm and $\Delta y=2.1\,$nm, where the radius was measured to be $r=1004.2\pm1.7\,$nm. The systematic errors resulting from the deviations of the shape to the parametrization (ideal circle) are difficult to quantify, since the precise extent of this deviation is unknown. However, the accuracy of the results apply to an ideal circular shape, which could be available in other experiments. Although this means a strict comparison in terms of accuracy per probe event is not possible, it can be concluded that a significant reduction in exposure can be achieved using the Bayes method instead of a linear sequence scan of the same area.

\onecolumngrid
\twocolumngrid

\section{Conclusion and Outlook}

We have shown how nanoscopic transmission microscopy benefits from the unique statistical properties of a single ion source in two different ways. On the one hand, the deterministic source in combination with a detector of finite quantum efficiency exhibits binomial noise characteristics which provides better SNR when compared to conventional Poissonian sources. This deterministic property additionally allows for imaging with negligible dark counts by gating the detector signal with respect to the extraction event. On the other hand, we demonstrated that the gain per particle in spatial information can be maximized by using the Bayesian experimental design method, in cases where additional topological information about the imaged structures is available. The high SNR and the optimization measures presented here are certainly not the only factors to determine how fast an image with a certain contrast can be acquired eventually, since this strongly depends on the repetition rate \textit{i.e.} the ion current. However, the approach at hand is suitable for applications where the acceptable current is limited, for example due to insulating, very pure or fragile samples which would otherwise suffer charging, contamination or damage from radiation.

To speed up the imaging procedure, a high power axial cooling beam or a precooling stage implemented by a magneto-optical trap could be employed. The reliability and also the resolution could be improved by integrating the source into a commercial ion beam column.

The temporal control of the ions down to the picosecond regime may enable ultrafast time resolved microscopy and stroboscopic measurements. Moreover, the time of flight information can be used to switch the focusing fields and thus circumvents the resolution-limiting Scherzer-theorem~\cite{schonhense2002correction}, which states that a rotationally symmetric ion optical lens with static electromagnetic fields, excluding space charges, always exhibits unavoidable spherical and chromatic aberrations. Through optical pumping it is feasible to implement a fully spin-polarized source \textit{e.g.} for sensing magnetic polarization of surfaces as pioneered in electron microscopy~\cite{duden1998spin}. Ultimately, the combination of control of the internal and external degrees of freedom of the ion would allow for the realization of matter wave interferometry with single ions~\cite{arndt2012focus,hasselbach2010progress}.

The apparatus was also conceived for deterministic ion implantation on the nanometer scale. This would enable the fabrication of scalable solid state quantum devices such as systems of coupled nitrogen vacancy color centres~\cite{dolde2013room}, coupled single phosphorous nuclear spins in silicon~\cite{kane1998silicon,jamieson2005controlled,pla2013high,veldhorst2015twoQubit} and cerium or praseodymium in yttrium orthosilicate~\cite{kolesov2012optical}. Here, imaging and implantation are highly complementary since absolute referencing, or more precisely alignment by imaging of the sample, is essential for an accurate positioning of dopants free of parallax errors.

\paragraph*{Acknowledgments}The authors acknowledge discussions with S. Prawer and G. Sch\"onhense. The team acknowledges financial support by the Volkswagen-Stiftung, the DFG-Forschergruppe (FOR 1493) and the EU-projects DIAMANT and SIQS (both FP7-ICT). FSK thanks for financial support from the DFG in the DIP program (FO 703/2-1). 

\bibliography{impbib}

\end{document}